\documentclass[twocolumn,letterpaper,showpacs,showkeys]{revtex4}
\usepackage{amsmath}
\usepackage{bm}

\begin{document}

\title{Resonance structures in the
multichannel quantum defect theory for the photofragmentation
processes involving one closed and many open channels} 

\author{Chun-Woo \surname{Lee}}
 \email{clee@madang.ajou.ac.kr}
\affiliation{Department of
Chemistry, Ajou 
University, Wonchun-Dong, Paldal-Gu, Suwon, 442-749, Korea.\\}

\date{\today}

\begin{abstract}
The transformation introduced by Giusti-Suzor and Fano and extended  
by Lecomte and Ueda for the study of
resonance structures in the multichannel quantum defect theory (MQDT)
is used to reformulate MQDT into  the forms 
having one-to-one  correspondence
with those in Fano's configuration mixing (CM) 
theory of resonance for the photofragmentation processes involving
one closed and many open channels.
The reformulation thus allows MQDT to have the full power 
of the CM  theory, still keeping its own strengths such as the
fundamental description of resonance phenomena without an
assumption of the presence of a discrete state as in CM. 

\end{abstract}

\maketitle

\section{Introduction}

Though multichannel quantum defect theory (MQDT) is a 
powerful theory of resonance capable of describing complex 
spectra including both bound and continuum regions with only a few 
parameters, resonance structures are not
transparently identified in its formulation because of the
indirect treatment of resonance\cite{Seaton83,Fanobook}. 
In order to identify resonance terms, one needs 
a special treatment like the one
Giusti-Suzor and Fano introduced for the
two channel case\cite{Suzor84}. They
noticed that 
the usual Lu-Fano 
plot often obscures  symmetry apparent in its
extended version.
The symmetry can be brought out in the MQDT
formulation by shifting the
origin of the plot to the center of symmetry
using the phase-shifted base pairs 
first considered in
Ref. \cite{Eissner69}:
\begin{equation}
(f,g) \rightarrow (f \cos \pi \mu -g \sin \pi \mu , \: g \cos \pi \mu
 + f \sin  \pi \mu ) .
\label{transform_base}
\end{equation}
By this phase renormalization, the diagonal elements of short-range
reactance matrices $K$ can be made zero 
so that    resonance structures are
separated from the background ones in two channel processes (Dubau and
Seaton also obtained the same results as Giusti-Suzor and Fano's ones
from a different approach\cite{DubauSeaton84}). 

Generalizations of their method to
the case involving more than two channels
have been done by Cooke and
Cromer\cite{Cooke85}, Lecomte\cite{Lecomte87}, Ueda\cite{Ueda87},
Giusti-Suzor and Lefebvre-Brion\cite{GiustiLefebvre84}, Wintgen and
Fridrich\cite{Wintgen87}, and Cohen\cite{Cohen98}. 
Lecomte  and Ueda  showed that, for
such a general 
case,  making the diagonal elements of reactance matrices
zero can only be achieved with an additional 
orthogonal transformation of basis
functions besides the phase renormalization\cite{Comment}.  Using
this transformation, Lecomte derived the best parameters 
for the description of total autoionization cross-sections 
shorn of the background part for the most general case 
involving many open and many closed channels.
Ueda derived total cross-section formulas
analogous to Fano's resonance formula for several cases 
including one closed and many open channels. 
Giusti-Suzor and
Lefebvre-Brion\cite{GiustiLefebvre84}, and Wintgen and
Friedrich\cite{Wintgen87} did the detailed studies
for the case of two closed and one open
channels and Cohen\cite{Cohen98}
involving two closed and two open channels.

One drawback of the above-mentioned work  is that partial 
cross-section formulas for photofragmentation processes were not 
dealt with. Recently, Lee\cite{Lee02_a} and Lee and Kim\cite{Lee02_b} 
derived the MQDT formulation which yielded 
the partial cross-section  formulas analogous to Fano's 
resonance formula and obtained 
the complete relation between MQDT
and the configuration mixing (CM) 
formulas\cite{Fano61,Fano65,Starace77,Lee95,
Fano78,Greene79,Mies79}.
But their work was restricted to the case involving only two open 
and one closed channels.
This paper extends their work to the case involving
many open channels and has succeeded in obtaining 
the same degree of results as the previous ones.

Section 2 describes the reformulation. Section 3 
derives the photofragmentation
cross-sections. Finally,
Section 4 gives the summary and discussion. 

\section{Reformulation}

In the multichannel quantum defect theory of photofragmentation
process, the coordinate $R$ for a relative motion of colliding
partners along which fragmentation takes place 
is divided into two ranges $R \le R_0$ and $R >R_0$, 
the inner and outer ones, respectively.  In contrast to the inner
range where transfers in energy, momentum, 
angular momentum, spin, or the formation of a transient complex
occur due to the 
strong interactions there, 
channels are decoupled in the outer range, and   the motion
is governed by ordinary second-order differential
equations and  described by superpositions of the
energy-normalized
regular and irregular base pair ($f_j (R)$,$g_j (R)$), or incoming and
outgoing base pair ($\exp (-ik_j R), \exp (ik_j R)$). 
For an $N$-channel system,
$N$ independent degenerate solutions of the Schr\"odinger equation 
for the decoupled motion in $R > R_0$ 
may be expressed into a standing-wave type 
\begin{equation}
\Psi_i (R, \omega ) = \sum_{j=1}^N \Phi_j (\omega ) 
[ f_j (R) \delta_{ji} -
g_j (R) K_{ji} ] , 
\label{Psi_stand}
\end{equation}
or an incoming-wave type
\begin{equation}
\Psi_i^{(-)} (R, \omega ) =  \sum_{j=1}^N
\Phi_j (\omega ) \left[ {\phi}_j^+ (R) \delta_{ji} -
{\phi}_j^- (R) S_{ji} \right], 
\label{Psi_in}
\end{equation}
where $\Phi_j (\omega )$ are the channel basis functions 
for the  coordinate space excluding $R$ and $\phi_j^{\pm}$ defined as 
$(\pm f_j +i g_j )/2$. 
$K_{ji}$ and $S_{ji}$ denote the ($j,i$)-elements of short-range
reactance and scattering matrices, respectively, and are related 
with each other in matrix notation by $S=(1-iK)(1+iK)^{-1}$
($S$ is here taken as a complex conjugate of the usual definition, 
for convenience).  
Using the quantum defect theory 
parameters $\eta_j$, $\beta_j$, and $D_j$ in Ref. 
\cite{Greene82} for an arbitrary field, $\phi_j^{\pm}$ are 
given in
the outer range $R > R_0$ by
$-i ({m_j}/2\pi k_j)^{1/2} \exp (\pm i \eta_j ) 
f_j^{\pm}$ for open channels and 
$\mp ({m_j}/{\pi \kappa_j})^{1/2} \exp (\pm i
\beta_j ) ( D_j f_j^+  \pm i D_j^{-1} f_j^- ) /2$ for closed 
channels, where $f_j^{\pm}$ denote $\exp (\pm ik_j R )$.  

Though all the $N$ solutions are needed to describe the motion 
in the intermediate range, some of them become closed and 
no longer exist in the limit of $R \rightarrow \infty$. 
In the present work, 
we will consider the case
involving only one closed and many, say $N_0$,
open channels  at large $R$, i.e. $N$ = $N_0
+1$.  
We will denote the set of open channels by $P$ and that of closed 
ones by $Q$. Open channels will be marked with 1,2,...,$N_0$ and 
the following single closed channel with $c$ instead of $N$ for 
easy recognition. 
Though meaningful only at large $R$,
still it may be convenient to keep 
the classification of channels 
as open or closed 
in the intermediate range. 
The wavefunction for the photofragmentation process 
into the $i$-th fragmentation channel, denoted as   
$\bm{\Psi}_i^{(-)}$, should satisfy the
incoming-wave boundary condition 
$\bm{\Psi}_i^{(-)}$ $\rightarrow$ $\sum_{j \in P}
\Phi_j \left( \phi_j^+ \delta_{ji} - \phi_j^- \bm{S}_{ji} \right) 
$ at large $R$\cite{S_eta} and can be obtained by making 
a linear combination of 
incoming channel basis functions $\Psi_i^{(-)}$ of 
Eq. (\ref{Psi_in}),
substituting the explicit forms for $\phi_j^{\pm}$ given
above and then setting the coefficients 
of exponentially rising terms to zero. 
This procedure yields
$\bm{{S}}$ =
${S}^{oo} - {S}^{oc}
({S}^{cc} 
- e^{2i {\beta} })^{-1} {S}^{co}$,
where the indices $o$ and $c$ stand for open and closed components, 
respectively. The second term 
shows that 
resonances come from the pole structure of the inverse matrix
$[S^{cc} -  \exp (2i \beta )]^{-1}$ due
to the closed channel. The first term $S^{oo}$, 
which contain couplings only among open channels,
cannot be regarded as corresponding to 
the background one
in the usual resonance theory such as the configuration
mixing  
method (CM) of Fano\cite{Fano61} because of its failure
to satisfy the unitary condition. 
To find the corresponding one to the background scattering matrix
$\bm{S_B}$ of CM, we rewrite 
the physical scattering matrix $\bm{S}$ 
into a form more analogous to that of CM as
\begin{equation}
\bm{S} = \sigma^{oo} + 2i \frac{(1+iK^{oo})^{-1} K^{oc}
K^{co} (1+iK^{oo})^{-1}} {\tan \beta +
\kappa^{cc} }  ,
\label{S_anal_CM}
\end{equation}
where $\kappa^{cc}$ is a new kind of complex reactance matrix 
studied extensively by Lecomte and defined by 
$S^{cc}$ = 
$(1-i \kappa^{cc} )(1+i \kappa^{cc})^{-1}$\cite{Lecomte87}.
The new scattering matrix  $\sigma^{oo}$ in Eq. (\ref{S_anal_CM}) is 
defined as $K^{oo}$ = 
$-i(1+\sigma^{oo})^{-1} (1-\sigma^{oo} )$ and, now pleasingly, 
unitary. From the definition, 
both symmetric $\sigma^{oo}$ and $K^{oo}$ are simultaneously
diagonalized as 
$U \exp (-2i \delta^{0} ) U^{T}$ and
$U\tan \delta^0 U^{T}$, respectively,  by the same
orthogonal matrix $U$. Eq. (\ref{S_anal_CM}) then becomes
\begin{equation}
\bm{S} = U e^{-i\delta^0}  \left( 1 + 2i \frac{\bm{\xi} \bm{\xi}^{T}}
{\tan \beta + \kappa^{cc}} 
\right) e^{-i\delta^0} U^{T} ,
\label{S_anal_CM2}
\end{equation}
where $\bm{\xi}$ denotes the column vector given by $\cos
\delta^0 U^{T} K^{oc}$. Notice that $\bm{\xi}^{T} \bm{\xi}$ = 
$K^{co}(1+{K^{oo}}^2 )^{-1} K^{oc}$ = $- \Im (\kappa^{cc} )$,
which is a scalar here, but generally a matrix and plays the key 
role in  Lecomte's work\cite{Lecomte87}.
Since $\bm{\xi}^{T} \bm{\xi}$ is   
positive definite, it can be denoted as $\bm{\xi}^{T} \bm{\xi}$ =
$\xi^2$. Elements of the column vector $\bm{\xi}$ are real but 
cannot be made positive, in general, by redefining $U$ 
since the latter is restricted
by $\det U =1$. The sum of their
squares is equal to $\xi^2$, i.e. $\sum_i \xi_i^2 = \xi^2$. 

In order to utilize Hazi's theorem that, 
for an isolated resonance in a multichannel system, 
sum of eigenphases satisfies the
resonance behavior of an elastic phase shift, 
the determinant of $\bm{S}$ is calculated 
by making use of the mathematical techniques in his
paper\cite{Hazi79} as
\begin{equation}
\det (\bm{S}) = e^{-2i \delta_{\Sigma}^0 } \left( \frac{\tan \beta +
{\kappa^{cc}}^* }{\tan \beta + \kappa^{cc} } \right) ,
\end{equation}
where $\delta_{\Sigma}^0$ denotes the sum of eigenphases 
of $\sigma^{oo}$, i.e.  
$\sum_j \delta_j^0$. If we let $\det (\bm{S} )$ = 
$\exp (-2i \delta_{\Sigma})$ with $\delta_{\Sigma} = 
\sum_j \delta_j$, then $\exp [-2i (\delta_{\Sigma} - 
\delta_{\Sigma}^0 )]$ = $(\tan \beta + {\kappa^{cc}}^* )/ 
(\tan \beta + \kappa^{cc} )$ and one obtains
\begin{equation}
\tan (\delta_{\Sigma} - \delta_{\Sigma}^0 ) \bigl[ \tan \beta + 
\Re (\kappa^{cc}) \bigr] = \Im (\kappa^{cc} ) = -\xi^2 .
\label{tan_delta_Sigma_orig}
\end{equation}
Following the lead of Giusti-Suzor and Fano\cite{Suzor84}, 
we may try to separate out  geometrical
factors from  channel coupling strength by translating  axes 
to make the Lu-Fano-like plot for $\delta_{\Sigma}$ vs. $\beta$
symmetrical by the phase renormalization described in 
Eq. (\ref{transform_base}). 
By the latter procedure, part of the  dynamics
manifested in the short-range reactance and scattering matrices $K$ 
and $S$ move into base pairs for motions in  decoupled channels. 
The net
effect is to transform the phase shifts $\eta_j$ ($j=1,...,N_0$) 
and $\beta$ of
the original base pairs for open and closed channels
into ${\eta_j}' = \eta_j + \pi \mu_j$ and 
${\beta}' = \beta + \pi \mu_c$, respectively. 
We will call the new 
representation,
in which the Lu-Fano-like plot for $\delta_{\Sigma}$ vs. $\beta$
is symmetrical, the tilde representation. 
The associated dynamical
parameters wii be accented by the tilde. Then
\begin{equation}
\tan \tilde{\delta}_{\Sigma} \tan \tilde{\beta} = \Im
(\tilde{\kappa}^{cc}) = - \tilde{\xi}^2 
\label{tan_delta_beta}
\end{equation}
with $\tilde{\delta}_{\Sigma}^0 =0$ and $\Re (\tilde{\kappa}^{cc} ) =
0 $. Eq. (\ref{tan_delta_beta}) implies that we can identify
$\tilde{\delta}_{\Sigma} $ with the phase shift $\delta_r$ due to 
the resonance. For  isolated resonances, $\delta_r$ varies 
as a function of energy as $\cot
\delta_r$ = $-\epsilon_r$ $\equiv$ $-2(E-E_0 )/ \Gamma$. 
Notice that 
Eq. (\ref{tan_delta_beta}) holds for all the resonances 
belonging to the same closed channel, 
yielding the extension of the definition of $\delta_r$ 
from $\cot \delta_r = - \epsilon_r$ to $\cot \delta_r$ = 
$- \tan \tilde{\beta}/\tilde{\xi}^2$. Here, we observe that 
there are infinite sets of \{$\mu_1 ,...,\mu_{N_0}$\} satisfying
$\tilde{\delta}_{\Sigma}^0$ = ${\delta}_{\Sigma}^0$ + 
$\pi \mu_{\Sigma}$ = 0 and thus  yielding
Eq. (\ref{tan_delta_beta}). A convenient choice may be $\mu_1$ =
$\mu_{\Sigma}$ and $\mu_j =0$ ($j=2,...,N_0$). Let us denote 
this particular set by $\mu_{S}^o$.
Observables are not affected by this arbitrariness 
as we will see later.

Now let us consider obtaining $\mu_{\Sigma}$ and $\mu_c$ 
which give rise to
the tilde representation.
The value of $\mu_c$ which yields $\Re (\tilde{\kappa}^{cc})=0$ is
easily obtained as 
$\tan 2 \pi \mu_c $ = $2\Re(\kappa ^{cc}
) /  
( 1- | \kappa^{cc}|^2 )$ 
from the transformation relation 
$\tilde{\kappa}^{cc}$ = $(\kappa^{cc} \sin \pi \mu_c + \cos \pi \mu_c
)^{-1} (\kappa^{cc} \cos \pi \mu_c - \sin \pi \mu_c )$
derived by Lecomte\cite{Lecomte87}. 
It may be expressed more compactly 
in terms of 
$S^{cc}$ as $\exp (-2i \pi \mu_c )$ = $S^{cc} /|S^{cc}|$,
indicating that the phase of $S^{cc}$ is removed so as to make 
$\tilde{S}^{cc}$ real and subsequently 
$\tilde{\kappa}^{cc}$ pure imaginary.
Next, let us consider obtaining
$\mu_{\Sigma}$ which yields ${\tilde{\delta}_{\Sigma}}^0 =0$. 
Under the phase renormalization, $\sigma^{oo}$ is transformed into
$\tilde{\sigma}^{oo}$ = $\exp (i\pi \mu^o )$\,
$\bigl\{ S^{oo}-S^{oc}S^{co}/[S^{cc}+\exp (-2i\pi \mu_c )]
\bigr\}$\,
$\exp (i\pi \mu^o )$. Then, the determinant of $\tilde{\sigma}^{oo}$ 
is calculated as $\exp (2i \pi \mu_{\Sigma})$  $[\det (S) + 
\exp (-2i\pi \mu_c ) \det (S^{oo}) ]/[S^{cc} + \exp (-2i\pi \mu_c )]$.
Since ${\tilde{\delta}_{\Sigma}}^0 =0$, $\det (\tilde{\sigma}^{oo})$
equals unity and one obtains the formula for $\exp (2i\pi
\mu_{\Sigma})$  as
$[S^{cc}+\exp (-2i\pi \mu_c )]/[\exp ( -2i\pi \mu_c ) \det (S^{oo}) 
+ \det (S)]$, where $\exp (-2i\pi \mu_c )$ = $S^{cc}/|S^{cc}|$ as 
already obtained. This formula for $\exp (2i\pi
\mu_{\Sigma})$ may be used to obtain the relation between 
$\tilde{\xi}^2$ and $\xi^2$ in conjunction with the relations
$\det (S)$ = 
$\exp [-2i\pi (\mu_{\Sigma}+\mu_c )]$ and $\det (S^{oo})$ = 
$\exp (-2i\pi \mu_{\Sigma} ) (1-\tilde{\xi}^2)/(1+\tilde{\xi}^2)$
available after studying the transformation
(\ref{K_trans_diagram}) later.
By substituting the relations
into the formula, we obtain
$\tilde{\xi}^2$ = $2\xi^2 /\{1+|\kappa^{cc}|^2+
[(1+|\kappa^{cc}|^2)^2 -4\xi^4 ]^{1/2} \}$.
It can be expressed
more compactly in terms of $|S^{cc}|$ as $\tilde{\xi}^2$ = 
$(1-|S^{cc}|)/(1+|S^{cc}|)$, indicating $\tanh \pi \alpha$ = 
$\tilde{\xi}^2$ if $S^{cc}$ is parameterized with 
Dubau and Seaton's complex quantum defect 
$\mu_c - i \alpha$ as 
$\exp [-2i\pi (\mu_c -i \alpha )]$\cite{DubauSeaton84}, 
which is equivalent to Eq. (35) 
of Ref. \cite{Suzor84}. Notice that $\tilde{\xi}^2 \le 1$. 
If $|\Im (\tilde{\kappa}^{cc})| >1$, Eq. (\ref{tan_delta_beta}) 
could be transformed into $\tan \tilde{\delta}_{\Sigma}'
\tan \tilde{\beta}'$ = $1/\Im
(\tilde{\kappa}^{cc})$ with 
$\tilde{\delta}_{\Sigma}$ = $\tilde{\delta}_{\Sigma}' +\pi /2$
and $\tilde{\beta}$ = $\tilde{\beta}' + \pi /2$ as described in 
Ref. \cite{Suzor84}. In this case, $\tilde{\xi}^2$ might be identified 
with $-1/\Im (\tilde{\kappa})$. 

In contrast to the two channel case\cite{Suzor84}, 
making the Lu-Fano-like plot
symmetrical is not enough to separate out the strength of 
channel coupling from the geometrical parameters in the short-range
reactance matrix, as evidenced by the nonzero $K^{oo}$ and $K^{cc}$. 
If there are more 
than two open channels, $K^{oo}$ cannot be made zero with
$\mu_{\Sigma}$ alone. The transformation to  make 
both $K^{oo}$ and
$K^{cc}$ zero was devised by Lecomte 
and Ueda\cite{Lecomte87,Ueda87} 
by extending the transformation of Giusti-Suzor and Fano.
In the present work, their prescription to make both 
reactance submatrices zero 
is a little modified in order to utilize the resonance structure 
in the sum of the eigenphase shifts as stated above. 
Let us briefly describe their transformation. It is
conveniently expressed in terms of $\phi_j^{\pm}$ as 
$\Phi_j {\phi_j '}^{\pm}$ = $\sum_i \Phi_i \phi_i^{\pm} W_{ij} 
\exp (\pm i\pi \mu_j )$,
where $W$ is an orthogonal
matrix with $W^{co}$ and $W^{oc}$  set to zero. Then, 
$W^{cc}$ is just
unity for the one closed channel case.
This leaves orthogonal transformations only
among base pairs of open channels. The second term 
$\exp (\pm i\pi \mu_j )$ induces the phases to be renormalized 
as ${\eta_j}'$ = $\eta_j + \pi \mu_j$ ($j=1,...,N_0$) for open
channels and as ${\beta}'$ = $\beta + \pi \mu_c$ for the closed
channel. 
The transformation is conveniently denoted by Lecomte as 
$T(\pi\mu^c , \pi\mu^o, W^{oo} )$, where $\mu^o$ is a set 
of $\mu_1 , ..., \mu_{N_o}$. For the transformation
composed of two successive operations like 
$T(0,\tilde{\delta}^{o},\tilde{U} ) T(0,0,U_r )$ of diagram 
(\ref{K_trans_diagram}) , consult Appendix \ref{Appendix_trans}.

Now let us go back to the problem of finding the transformation 
which makes $\tilde{K}^{oo}$ 
and $\tilde{K}^{cc}$ zero. 
Though this problem is already solved by Lecomte\cite{Lecomte87}, 
let us
give a brief description of it for the subsequent
description.  With $T(0, \pi\mu^o,W^{oo})$, $\tilde{K}^{oo}$ 
is transformed
into ${\tilde{K'}}^{oo}$ = 
$( \tilde{K}_W \sin \pi \mu^o  + 
\cos \pi \mu^o )^{-1} (\tilde{K}_W \cos \pi \mu^o 
-\sin \pi \mu^o )$ where 
$\tilde{K}_W$ = ${W^{oo}}^{T} {\tilde{K}}^{oo} W^{oo}$. 
Let $\tilde{U}$ diagonalize ${\tilde{K}}^{oo}$, i.e.,
${\tilde{U}}^{T} {\tilde{K}}^{oo}\tilde{U}$ = 
$\tan {\tilde{\delta}}^0$. 
Then, $T(0,{\tilde{\delta}}^0 , 
\tilde{U})$ transforms $\tilde{K}^{oo}$ into a zero matrix.
${\tilde{K}}^{cc}$ is transformed into zero too as a by-product,
which derives from two theorems. First, ${\tilde{\kappa}}^{cc}$ does 
not change value under any transformation with $\mu_c =0$.
Therefore, $\Re (\tilde{\kappa}^{cc})$ = $0$ 
remains invariant under $T(0,{\tilde{\delta}}^0 , 
\tilde{U})$.
Secondly, $\kappa^{cc}$ = $K^{cc} - i K^{co}K^{oc}$ 
if $K^{oo}=0$, whereby one has $K^{cc}$ = 
$\Re (\kappa^{cc})$.

Let us call the new representation generated by 
$T(0,{\tilde{\delta}}^0 , \tilde{U})$
the bar-representation. In this case, 
the physical scattering matrix $\bar{\bm{S}}$ becomes  
$1$ $-$  
$2i \exp ( -i \delta_r ) \sin \delta_r \tilde{\bm{\xi}}$
$\tilde{\bm{\xi}}^{T}/{\tilde{\xi}^2}$. 
Since $\tilde{\bm{\xi}}\tilde{\bm{\xi}}^{T}$ is a 
$N_0 \times N_0$ symmetric matrix 
of rank 1, it can be diagonalized by some orthogonal matrix, say
$U_r$, as 
\begin{equation}
U_r \tilde{\bm{\xi}}\tilde{\bm{\xi}}^{T} U_r^{T} = \tilde{\xi}^2 
\begin{pmatrix}
1&0&\dots&0\\
0&0&\dots&0\\
\vdots&\vdots&\ddots&\vdots\\
0&0&\cdots&0 
\end{pmatrix}  
\equiv \tilde{\xi}^2 \bm{p_r} ,
\label{projec}
\end{equation}
where $\bm{p_r}$ satisfies the property of a projection matrix. 
If we put $U_r^{T} \bm{p_r} U_r$ = $\bm{P_r}$,  
$\bar{\bm{S}}$ can be written as $\exp (-2i\delta_r \bm{P_r})$,
which suggests a new representation where 
the physical scattering matrix
is diagonal as $\exp (-2i \delta_r \bm{p_r} )$.
It is easily seen that the new representation is generated 
by $T(0,0,U_r )$. 
Let us call the new representation the r-representation. 
In this representation, the
short-range $N \times N$ reactance matrix $K_r$ has only 
two nonzero elements whose value is just 
the strength of channel coupling:
\begin{equation}
K_r = 
\begin{pmatrix}
0&0&\dots&\xi\\
0&0&\dots&0\\
\vdots&\vdots&\ddots&\vdots\\
\xi&0&\cdots&0  
\end{pmatrix}  .
\label{K_r}
\end{equation}
With this $K_r$, only ${\Psi_r}_1$ and 
${\Psi_r}_c$ have coupling terms (recall that $c$ is actually $N$).
${\Psi_r}_1$ is dubbed
the `effective continuum'  by others\cite{Cooke85} 
and corresponds to 
Fano's `a' state $\psi_E^{(a)}$\cite{Fano65}. 
$\bar{\bm{S}}$ and $\bm{S_r}$ only contain the resonant dynamics 
and may be expressed as 
$\exp (-2i \delta_r \bar{\kappa}^{oo} / \bar{\kappa}^{cc} )$ 
and 
$\exp (-2i \delta_r {\kappa}_r^{oo} / {\kappa}_r^{cc} )$,
respectively. 
The process described so far can be 
summarized in the  following diagram:
\begin{eqnarray}
&\left\{ 
\begin{array}{l}K\\ \delta_{\Sigma}^0 \ne 0 
\\ \Re(\kappa^{cc})\ne 0 
\\ \Im(\kappa^{cc})=- \xi^2
\\ \Re(\kappa^{oo})\ne 0
\\ \beta
\end{array} 
\right. 
\xrightarrow{T \bigl(\pi\mu_c ,\pi\mu_S^o , I^{oo} 
\bigr)}
\left\{ 
\begin{array}{l}\tilde{K}\\ \tilde{\delta}_{\Sigma}^0 = 0
~(\tilde{\sigma}^{oo} =1)
\\ \Re(\tilde{\kappa}^{cc})=0 
\\ \Im(\tilde{\kappa}^{cc})=- \tilde{\xi}^2
\\ \Re(\tilde{\kappa}^{oo}) \ne 0
\\ \tilde{\beta} = \beta + \pi \mu_c
\end{array} 
\right. 
\nonumber\\
&\xrightarrow{T(0,\tilde{\delta}^0 ,\tilde{U})}
\left\{
\begin{array}{l} \Bar{K}\\ \bar{\delta}_{\Sigma}^0 = 0
~(\Bar{\sigma}^{oo} =1)
\\\Re(\Bar{\kappa}^{cc})=0 
\\ \Im(\Bar{\kappa}^{cc})=- \tilde{\xi}^2
\\ \Re(\Bar{\kappa}^{oo})= 0 ,~ 
   \Im(\Bar{\kappa}^{oo})=- \tilde{\xi}^2 \bm{P_r}
\\ \bar{\beta} = \tilde{\beta}
\end{array} \right.  
\nonumber \\
&\xrightarrow{T(0 ,0,U_r)}
\left\{
\begin{array}{l}K_r\\ {{\delta_r}^0_{\Sigma}} = 0
~({\sigma}_r^{oo} =1)
\\ \Re(\kappa_r^{cc})=0
\\ \Im(\kappa_r^{cc})=- \tilde{\xi}^2
\\ \Re(\kappa_r^{oo})= 0, ~
   \Im({\kappa}_r^{oo})=- \tilde{\xi}^2 \bm{p_r}
\\ \beta_r = \tilde{\beta}
\end{array} \right.  ,
\label{K_trans_diagram}
\end{eqnarray}
where the set $\mu_S^o$ is given by$\{ \mu_{\Sigma},0,...,0 \}$
as introduced before.
Once made symmetrical by the translation of the axes, 
the graph 
of $\delta_{\Sigma}$ vs. $\tilde{\beta}$ remains moveless
under further transformations
$T(0,\tilde{\delta}^0 ,\tilde{U})$ and
$T(0 ,0,U_r)$. We will call this kind of 
representations, which share the identical location 
of the graph on the plot, 
the resonance-centered 
representation hereinafter.  
The departure of our procedure 
from Lecomte's one lies in that 
$K^{oo}$ is not made into a zero matrix 
as the first step but only 
phases are
renormalized  so as to symmetrize  the plot for $\delta_{\Sigma}$
vs. $\tilde{\beta}$. 
This offers several
advantages. 
The background part  eliminated in the 
last two representations of diagram 
(\ref{K_trans_diagram}) is nothing but disentangled 
by the first process, 
enabling us to identify its MQDT form.
The fact that the last two transformations 
operate only on the open channel set indicates 
that the resonance effect due to the closed channel is already
fully accounted for by the phase renormalization alone.
This enables us to separate out the background and resonant
contributions in the partial cross-section formulas, which 
is the topic of the next section.

\pagebreak

\section{Photofragmentation cross-section formulas}
\label{sec:cross_section}

Let us  consider the photofragmentation processes from an initial
bound state to the $j$-th fragment one. The fragment state may be 
described by an incoming-wave as follows\cite{Lecomte87,Lee02_b} 
\begin{equation}
\bm{{\Psi }}_j^{(-)} = {\Psi }_j^{(-)} +  
{\Psi }_c^{(-)} \biggl[ \frac{\tan
{\beta }+i }{\tan {\beta } + {\kappa }^{cc} } 
 {K}^{co} (-i+ {K}^{oo} )^{-1}\biggr]_{cj} .
\label{Psi_prime_asym_re}
\end{equation}
Notice that the term $(\tan
\beta  +i)(\tan \beta  + {\kappa }^{cc})^{-1}$  is the same
for all the resonance-centered representations as $(\tan
\beta  +i)[\tan \beta  + i\Im ({\kappa }^{cc})]^{-1}$ 
and is very energy-sensitive as can be seen from its another
expression $-(i/\tilde{\xi}) (d\delta_r /d
\tilde{\beta})^{1/2} \exp [-i(\tilde{\beta}+\delta_r )]$ 
obtainable from it by means of
$\tan \tilde{\beta} \tan \delta_r $ = $-\tilde{\xi}^2$. 
Let us introduce new short-range wavefunctions
${M}_j^{(-)}$ and 
${N}_j^{(-)}$ defined only for open channels by
\begin{eqnarray}
{M}_j^{(-)} &=& {\Psi}_j^{(-)} +  {\Psi}_c^{(-)}
\bigl[ {K}^{co}(-i+ {K}^{oo})^{-1} \bigr]_{cj} ,
\nonumber \\
{N}_j^{(-)} &=& {\Psi}_j^{(-)} +\tfrac{i}{\kappa^{cc}}
{\Psi}_c^{(-)} 
\bigl[ \tilde{K}^{co}(-i+ \tilde{K}^{oo})^{-1} \bigr]_{cj} .
\label{tilde_M_N}
\end{eqnarray}
Using these wavefunctions, the square of the modulus of 
the transition dipole moment 
${\bm{D}}_j^{(-)}$ [$\equiv ({\bm{\Psi}}_j^{(-)}|T|i )$]
may be expressed 
as
\begin{equation}
\Bigl| {\bm{D}}_j^{(-)} \Bigr| ^2 = 
\Bigl| \bigl({M}_j^{(-)}|T|i\bigr)\Bigr| ^2 
\frac{| [\tan {\beta}+\Re (\kappa^{cc})] / {\xi}^2 +
{q}_j |^2}{[\tan  {\beta} +\Re (\kappa^{cc})]^2/ {\xi}^4 +1 }   ,
\label{tran_dipol}
\end{equation}
where $T$ is the dipole moment operator, $i$ stands for the initial
bound state, and the complex line profile index parameter 
${q}_j$ is given by ${q}_j$ = $i
\bigl({N}_j^{(-)}|T|i\bigr)/ 
\bigl({M}_j^{(-)}|T|i\bigr)$. 
If the tilde representation is considered,
the relations $\Re (\tilde{\kappa}^{cc} )=0$ and 
$\tan \tilde{\beta} \tan \delta_r $ = $-\tilde{\xi}^2$
holding for it may  be used 
to put Eq. (\ref{tran_dipol})  into a Beutler-Fano form:
\begin{equation}
\Bigl| \tilde{\bm{D}}_j^{(-)} \Bigr| ^2 = 
\Bigl| \bigl(\tilde{M}_j^{(-)}|T|i\bigr)\Bigr| ^2 
\frac{| -\cot \delta_r +
\tilde{q}_j |^2}{ \cot ^2 \delta_r + 1 }   .
\label{tilde_tran_dipol}
\end{equation}
$| \tilde{\bm{D}}_j^{(-)} |$ equals 
$| {\bm{D}}_j^{(-)} |$ since 
$\tilde{\bm{{\Psi }}}_j^{(-)}$ differs from 
$\bm{{\Psi }}_j^{(-)}$ by $
\exp (i\pi \mu_j )$\cite{Lecomte87,Lee02_b}. 
Phase renormalization does not change the 
absolute magnitude of a transition dipole matrix element 
but is instrumental in making 
the transition dipole matrix element into a Beutler-Fano form
since the latter is only obtained in the tilde representation, 
or in the one obtainable from the tilde representation 
by the phase renormalization 
which keeps the eigenphase sum of physical scattering matrix
unchanged. 

For the resonance-centered representations, the physical incoming
wavefunctions  may be 
expressed as 
$\bm{\Psi}_j^{(-)}$ = 
$e^{-i\delta_r } \bigl( M_j^{(-)} \cos \delta_r + i N_j^{(-)} \sin
\delta_r \bigr)$,
which shows that 
$M_j^{(-)}$ is the sole contributor to  
the physical incoming-waves at the energy 
where the phase shift $\delta_r$ due to 
the resonance is zero.  
Its comparison 
with CM's physical incoming wavefunction\cite{Lee00}
\begin{widetext}
\begin{equation}
\bm{\Psi}_E^{-(j)} (\mathrm{CM}) = e^{-i\delta_r } 
\Biggl\{ \psi_E^{-(j)} \cos \delta_r + i \biggl[ 
\Bigl( 1-  |\psi_E^{(a)}\rangle \langle \psi_E^{(a)}| +
\frac{i\Phi}{\pi (\sum_k |V_{kE}|^2)^{1/2}} 
\langle \psi_E^{(a)}| \Bigr) \psi_E^{-(j)}\biggr] \sin \delta_r 
\Biggr\} 
\label{Psi_j_CM}
\end{equation}
\end{widetext}
suggests a one-to-one correspondence between  
${M}_j^{(-)}$ and $\psi_E^{-(j)}$
and also between $N_j^{(-)}$  and
the term inside the square brackets which constitutes
the second term 
inside the curly braces of the 
right-hand side of Eq. (\ref{Psi_j_CM}). 
The one-to-one correspondence between ${M}_j^{(-)}$ and
$\psi_E^{-(j)}$ can also be seen in the 
asymptotic forms: the open channel part of 
the  decoupled form
$\sum_{i \in P} \Phi_i ( \phi_i^+ \delta_{ij} - \phi_i^- 
\sigma_{ij}^{oo})$ $-$ $\Phi_c (\phi_c^+ + \phi_c^-)
[(1+S^{cc})^{-1}S^{co}]_{cj}$ of ${M}_j^{(-)}$ 
( $R \ge R_0$) is identical to the 
asymptotic form of 
$\psi_E^{-(j)}$ if the one-to-one correspondence between
$\sigma^{oo}$ and $\bm{S_B}$ is taken into account.
The decoupled form, however, contains an additional 
closed channel term which rises  exponentially  
at large $R$,
showing that ${M}_j^{(-)}$ by itself is not a physically 
acceptable wavefunction in contrast to the background one.
But its contribution to cross-sections is still finite
since it is multiplied by an initial
wavefunction that may be reasonably assumed to be bound.
This indicates that the background part $\bm{S_B}$ of the
scattering matrix $\bm{S}$ (=$\bm{S_B S_R}$) of CM actually
contains  closed channel contributions.  
The closed channel contribution into
${{M}}_j^{(-)}$ is given by the form 
of $\phi_c^+ + \phi_c^-$ which is 
equal to $i$ times the irregular function $g_i$. 
It shows that the regular function for the closed channel
contributes nothing to ${M}_j^{(-)}$, 
presumably indicating that  ${M}_j^{(-)}$ is
the form of minimal closed channel contribution
in the intermediate and  reaction zones
and thus in the observables.
This claim requires further study for sure.

Eq. (\ref{tilde_tran_dipol}) may be used to obtain $\bigl|
(\tilde{M}_j^{(-)}|T|i )\bigr| ^2$, $\Re (\tilde{q_j})$, $\Im
(\tilde{q_j})$, $\mu_c$, and $\tilde{\xi}^2$ 
from the experimental data using the method
developed in the field of modeling of data\cite{Recipes}
(the form of $\beta$ as a function of energy 
needed for data fitting is given analytically for most fields 
but should be obtained numerically for the zero field 
using the Milne procedure described in Ref. \cite{Greene82}).
But Eq. (\ref{tilde_tran_dipol}) is not expressed 
in terms of parameters whose physical origins are clearly 
identified.
The r-representation may be used for that purpose
since the  channel coupling strength
can only be  completely disentangled there  from the geometrical
factors  and 
the formulation is additionally simplified  too
by the fact that only one open process can be involved 
there for
the resonance.
Introducing the r-representation is equivalent to 
visualizing  the
photofragmentation process 
as being excited to eigenchannels of $\bm{S_r}$ 
but observed in the detector through their
projections to the detector eigenchannels.
By the fact that $\bm{S_r}$ is
already diagonalized as $\exp (-2i \delta_r \bm{p_r})$ 
with eigenvalues \{$\exp (-2i\delta_r ),1,...,1$\} and $K_r^{cc} =0$,
the well-known formulas for the eigenchannel
wavefunctions ${\bm{\Psi_r}}_k^{(\textrm{eig})}$ given
as the superpositions of standing-waves\cite{Lee02_b}, 
i.e. ${\Psi_r}_k \cos \delta_k + {\Psi_r}_c {Z_r}_{ck} 
\cos \tilde{\beta}_c$
with ${Z_r}_{ck} \cos \tilde{\beta}_c $ = 
$-( \tan  \tilde{\beta}_c )^{-1}
(K_r^{co}\cos \delta )_{ck}$, become reduced to
${\Psi_r}_1 \cos \delta_r + {\Psi_r}_c \sin
\delta_r / \tilde{\xi}$ for $k$ = 1 and  ${\Psi_r}_k$ otherwise. 
The transition dipole moments to 
${\bm{\Psi_r}}_i^{(\textrm{eig})}$ can, then,
be obtained as
\begin{eqnarray}
{\bm{D_r}}_1^{(\textrm{eig})}  &=&
-({\Psi_r}_1|T|i) \frac{ \tan \tilde{\beta} / \tilde{\xi}^2 +
q_r }{\bigl( \tan ^2 \tilde{\beta} / 
\tilde{\xi}^4 +1 \bigr)^{1/2}}   , 
\nonumber\\
{\bm{D_r}}_k^{(\textrm{eig})}  &=&
({\Psi_r}_k|T|i),~~~\text{($k \ne 1$)} 
\label{tran_dipol_r}
\end{eqnarray}
with the line profile index $q_r$ defined as 
$q_r$ = $-
({\Psi_r}_c|T|i)/[\xi({\Psi_r}_1|T|i)]$,
which is clearly real because 
the standing waves ${\Psi_r}_1$
and ${\Psi_r}_c$ are real\cite{Convention}. 
From the unitary relation between $\bm{\Psi}_j^{(-)}$ and 
${\bm{\Psi_r}}_j$, we have 
$\sum_{j \in P} \bigl|\tilde{\bm{D}}_j^{(-)}\bigr|^2$ =
$\sum_{k \in P}
\bigl|{\bm{D_r}}_k^{(\textrm{eig})} \bigr|^2$. Using this relation,
Eq. (\ref{tilde_tran_dipol}) becomes 
\begin{eqnarray}
\sum_{j \in P} \bigl|\tilde{\bm{D}}_j^{(-)}\bigr|^2 
=&  \bigl| ({\Psi_r}_1|T|i) \bigr|^2 \frac{ ( \tan \tilde{\beta} /
\tilde{\xi}^2 + q_r )^2 }{\tan ^2 \tilde{\beta} 
/ \tilde{\xi}^4 +1 } \nonumber \\ &+
{\sum_{k \in P}}' {\bigl| ({\Psi_r}_k|T|i)\bigr|^2 },
\label{sum_D_sqr}
\end{eqnarray}
where the prime on the summation symbol
denotes  that $k=1$ is excluded in the summation.
Eq. (\ref{sum_D_sqr}) directly corresponds to 
the well-known total cross-section
formula $\sigma_{\mathrm{tot}}$ = $\sigma_a (\epsilon + q)^2 / 
(\epsilon^2 +1 ) + \sigma_b$
of CM for photofragmentation 
in the neighborhood of an isolated
resonance if the one-to-one correspondence 
between ${\Psi_r}_1$ and $\psi_E^{(a)}$, described below, is taken
into account\cite{Fano65}.

Since $\tilde{\bm{\Psi}}_j^{(-)}$ and 
${\bm{\Psi_r}}^{(\textrm{eig})}_k$ are
energy-normalized and related by a unitary transformation, 
their transition dipole moments are also  related 
by the same unitary transformation as 
$\tilde{\bm{D}}_j^{(-)}$ = 
$\sum_{k\in P} {\bm{D_r}}_k^{(\textrm{eig})}  
\bigl(\tilde{\bm{\Psi}}_j^{(-)}|{\bm{\Psi_r}}^{(\textrm{eig})}_k \bigr)$.
Using the transformation relation
$\bigl(\tilde{\bm{\Psi}}_j^{(-)}|{\bm{\Psi_r}}^{(\textrm{eig})}_1\bigr)$
= $\exp (i \delta_r ) ({\tilde{M}}_j^{(-)}| {\Psi_r}_1 )$ and 
$\bigl(\tilde{\bm{\Psi}}_j^{(-)}|{\bm{\Psi_r}}^{(\textrm{eig})}_k\bigr)$
= $({\tilde{M}}_j^{(-)}| {\Psi_r}_k )$ ($k \ne 1$)  derived in
Appendix \ref{Appendix_trans} and 
the formulas for ${\bm{D_r}}_k^{(\textrm{eig})}$ 
given in Eq. (\ref{tran_dipol_r}), 
the transition dipole moment
to the $j$-th fragmentation 
channel can be obtained as
\begin{eqnarray}
\tilde{\bm{D}}_j^{(-)} &= 
\bigl(\tilde{M}_j^{(-)}|T|i\bigr) \Biggl[
 \frac{ \tan \tilde{\beta} / \tilde{\xi}^2 +
q_r }{\tan \tilde{\beta} / \tilde{\xi}^2 +i }
\frac{\bigl(\tilde{M}_j^{(-)}|{\Psi_r }_1\bigr)
\bigl({\Psi_r}_1|T|i\bigr)}{\bigl(\tilde{M}_j^{(-)}|T|i\bigr)} 
\nonumber \\
&+  {\sum_{k \in P}}' \frac{\bigl(\tilde{M}_j^{(-)}|{\Psi_r}_k\bigr)
\bigl({\Psi_r}_k|T|i\bigr) 
}{\bigl(\tilde{M}_j^{(-)}|T|i\bigr)} 
\Biggr] . 
\label{tran_dipol_2}
\end{eqnarray}
Let us define $\tilde{\rho}_j$ as
\begin{equation}
\tilde{\rho}_j = \frac{\bigl(\tilde{M}_j^{(-)}|{\Psi_r}_1\bigr)
\bigl({\Psi_r}_1|T|i\bigr)} {\bigl(\tilde{M}_j^{(-)}| T
|i\bigr)}  = \frac{\bigl(P_{r1}\tilde{M}_j^{(-)}|T|i\bigr)} 
{\bigl(\tilde{M}_j^{(-)}| T |i\bigr)}  
\label{rho_j}
\end{equation}
with $P_{r1} = |{\Psi_r}_1 \rangle \langle {\Psi_r}_1 |$ 
in analogous to $\rho_j$ of CM (identical to Starace's $\alpha^*
(jE)$\cite{Starace77}) defined as
$(P_a \psi_E^{-(j)}|T|i)/(\psi_E^{-(j)}|T|i)$ with $P_a$ = 
$|\psi_E^{(a)} \rangle \langle \psi_E^{(a)} |$\cite{Lee95}. 
(Notice that all the representations connected by the phase 
renormalization have the common value of $\tilde{\rho}_j$.
This is consistent with the fact that phase 
renormalization does not change the absolute magnitude of 
a transition dipole matrix element.)
Then it may be shown that the second term inside the brackets of the
right-hand side of Eq. (\ref{tran_dipol_2}) is just 
$1-\tilde{\rho}_j$.
Substituting this and Eqs. (\ref{rho_j}) into 
Eq. (\ref{tran_dipol_2}), we  obtain Eq. (\ref{tilde_tran_dipol}) but
now with $\tilde{q}_j$ expressed in terms of parameters 
$q_r$ and $\tilde{\rho}_j$ of clear physical origin
as 
\begin{equation}
\tilde{q}_j = q_r \tilde{\rho}_j +i (1-\tilde{\rho}_j) .
\end{equation}

Before, we claimed that 
not $\tilde{\Psi}_j^{(-)}$ but $\tilde{M}_j^{(-)}$ 
corresponds to 
the background wavefunction $\psi_E^{-(j)}$ of CM. 
A similar correspondence  may be claimed for ${M_r}_i^{(-)}$. 
But, for the r-representation, 
${M_r}_i^{(-)}$  equals  ${\Psi_r}_i$, as shown in 
Appendix \ref{Appendix_trans}. Therefore, we claim that
${\Psi_r}_i$ corresponds to 
CM's $\psi_E^{(a)}$ for $i$=1 and $\psi_E^{( \lambda )}$ 
($\lambda \ne a$) otherwise. Notice that ${\Psi_r}_i$ are
real quantities as are $\psi_E^{( \lambda )}$, 
which is the reason why ${\Psi_r}_i$ is used preferably to 
${M_r}_i^{(-)}$ in the above equations.  
The claim is bolstered by the 
same one-to-one correspondence between wavefunctions 
found from the comparison of 
MQDT's $\tilde{\rho}_j$ with CM's $\rho_j$.
Here, we only talked about the analogy between 
formulas of two theories not the actual 
relations of corresponding terms in two theories.
The relations may be derivable from the prescription described in 
Ref. \cite{Lee02_b}. For example, it may be found that
${\Psi_r}_i$ equals $\psi_E^{(a)}$ + $i\tilde{\xi} \Phi_c (\phi_c^+ +
\phi_c^- )$ + $O(\tilde{\xi}^2)$ 
for $i=1$ and $\psi_E^{( \lambda )}$ + $O(\tilde{\xi}^2)$
($\lambda \ne a$)  otherwise.

Finally,  
let us consider about dynamical parameters  
extractable from
the total and partial photofragmentation cross-sections.
Since total cross-sections are proportional to 
$\sum_{j \in P} \bigl|\tilde{\bm{D}}_j^{(-)}\bigr|^2$, 
Eq. (\ref{sum_D_sqr}) may be used to fit the experimental data of
total cross-sections. 
Levenberg-Marquardt method\cite{Recipes} may
be employed for such a
data fitting to obtain the information on $\bigl| ({\Psi_r}_1|T|i)
\bigr|^2$, ${\sum_{j \in P}}' {\bigl| ({\Psi_r}_j|T|i)\bigr|^2 }$,
$q_r$, $\tilde{\xi}^2$, and $\mu_c$.   
Information on the absolute value of $({\Psi_r}_c|T|i)$ 
and its relative sign to $\tilde{\xi}
({\Psi_r}_1|T|i)$ may be obtained from $q_r$, 
since it is defined as $- ({\Psi_r}_c|T|i)/[\tilde{\xi}
({\Psi_r}_1|T|i) ]$.
For the partial cross-sections,
$| \tan \tilde{\beta} / \tilde{\xi}^2 +
\tilde{q}_j |^2 /(\tan ^2 \tilde{\beta} / 
\tilde{\xi}^4 +1 )$ is changed to the form 
consisted of real terms as
$[ \tan \tilde{\beta} / \tilde{\xi}^2 +\Re (q_j )]^2/
(\tan ^2 \tilde{\beta} / \tilde{\xi}^4 +1 )$ 
+ $[\Im (q_j )]^2/
(\tan ^2 \tilde{\beta} / \tilde{\xi}^4 +1 )$, 
which may be used to extract $\Re ( \tilde{q}_j )$ and 
$[\Im ( \tilde{q}_j )]^2$. 
Notice that  the data fitting leaves the sign of $\Im 
(\tilde{q}_j )$  undetermined. 
After $\tilde{q}_j$ is obtained up to the sign 
of its imaginary part, 
$\tilde{\rho}_j$  is obtainable
from  the relation 
$\tilde{q}_j$ = 
$q_r \tilde{\rho}_j +i (1-\tilde{\rho}_j)$, which yields the 
quadratic equation for $\tilde{\rho}_j$ and eventually 
gives two $\tilde{\rho}_j$ compatible with both 
$\tilde{q}_j$ and 
${\tilde{q}}_j^*$. 
From $\tilde{\rho}_j$, 
the information is obtained on the projection factor 
$\bigl(\tilde{M}_j^{(-)}|{\Psi_r}_1\bigr)$ 
since other factors like
$\bigl({\Psi_r}_1|T|i\bigr)$ and the absolute magnitude of
$\bigl(\tilde{M}_j^{(-)}|T|i\bigr)$ constituting
$\tilde{\rho}_j$ are already obtained.
The projection factor is related to the  component of
$\tilde{\bm{\xi}}$ but not directly because
the latter pertains to the eigenchannels of
$\tilde{\sigma}^{oo}$. The relation is 
given by 
$|\tilde{\xi}_k /\tilde{\xi}|$ 
=
$|\sum_{j \in P} \tilde{U}_{kj}^{T} 
\bigl(\tilde{M}_j^{(-)}|{\Psi_r}_1\bigr)|$
where absolute value is taken to get rid of an 
unimportant phase factor.

\section{Summary and discussion}

We confirmed again, for the case involving one closed
and many open channels, 
the striking similarities between MQDT and CM formulas 
found for the case involving one closed 
and two open channels\cite{Lee02_b} 
if  MQDT is reformulated by means of
Giusti-Suzor and Fano's 
phase renormalization and Lecomte and Ueda's additional 
orthogonal transformation. 
The unitarity of $\tilde{\sigma}^{oo}$ 
[$\equiv (1-i\tilde{K}^{oo})(1+i\tilde{K}^{oo})^{-1}$]
and its simultaneous diagonalizability with $\tilde{K}^{oo}$
by the same orthogonal transformation are newly found 
to play the pivotal 
role in the reformulation.
By this reformulation, we found the one-to-one 
correspondence between 
two different manifestations, $\tilde{M}_j^{(-)}$ 
and ${\Psi_r}_j$, of 
the form ${\Psi}_j^{(-)} +  {\Psi}_c^{(-)}
\bigl[ {K}^{co}(-i+ {K}^{oo})^{-1} \bigr]_{cj}$ of MQDT
and the background wavefunction  
$\psi_E^{-(j)}$ and Fano's `abc..' states of CM, respectively,   
and also between $\tilde{\sigma}^{oo}$ of MQDT
and the background scattering matrix 
$\bm{S_B}$ of CM.
Under this correspondence,
formulas in both theories 
exactly coincide with each other when 
further one-to-one correspondence coming from
the extension of $-\cot \delta_r$ = $2(E-E_0 )/\Gamma$ of CM to
$\tan \delta_r \tan \tilde{\beta}$ = $-\tilde{\xi}^2$ of MQDT
taken into account. 
Note that the  reformulation allows MQDT to have the full power 
of the CM  theory, still keeping its own strengths such as the 
fundamental description of resonance phenomenon without any
assumption of the presence of a discrete state as in CM.

\begin{acknowledgments}
I am greatly thankful to Ji-Hyun Kim for his help in the 
first stage of the work. 
This work was supported
by KRF under contract No. 99-041-D00251 D3001.
\end{acknowledgments}


\appendix

\section{Transformation relations among various wavefunctions}
\label{Appendix_trans}

Here, we want to prove the relations
$\bigl(\tilde{\bm{\Psi}}_j^{(-)}|
{\bm{\Psi_r}}^{(\textrm{eig})}_1\bigr)$ 
= $\exp (i \delta_r ) ({\tilde{M}}_j^{(-)}| {\Psi_r}_1 )$ and 
$\bigl(\tilde{\bm{\Psi}}_j^{(-)}|
{\bm{\Psi_r}}^{(\textrm{eig})}_i\bigr)$ 
= $({\tilde{M}}_j^{(-)}| {\Psi_r}_i )$ ($i \ne 1$),
which can be recast as 
$\bigl(\tilde{\bm{\Psi}}_j^{(-)}|
{\bm{\Psi_r}}^{(-)}_i\bigr)$ 
= $({\tilde{M}}_j^{(-)}| {\Psi_r}_i )$ 
by means of the relation 
between eigenchannels
and incoming-waves in the r-representation given by
${\bm{\Psi_r}}_i^{(\textrm{eig})}$ 
= ${\bm{\Psi_r}}_i^{(-)} \exp(i \delta_r )$ for 
$i$ = 1 and equals otherwise. 
Let us consider the transformation from the tilde 
representation into the r-one. It is performed by 
two transformations $T(0, \tilde{\delta}^0 , \tilde{U})$ and 
$T(0,0,U_r )$. If we denote the $N_0 \times N_0$ unitary
transformation  
$\tilde{U} \exp (i\tilde{\delta}^0 ) U_r$ as $V$,
then, from the definition of the transformation extended by Lecomte
and Ueda,  the following transformation 
relations are obtained: 
$\Phi_j {\phi_r}_j^+$ = $\sum_{i \in P} 
\Phi_i \tilde{\phi}_i^+ V_{ij}$
and 
$\Phi_j {\phi_r}_j^-$ = $\sum_{i \in P} 
\Phi_i \tilde{\phi}_i^- V_{ij}^*$
for $j \in P$ and $\Phi_c {\phi_r}_c^{\pm}$ = 
$\Phi_c \tilde{\phi}_c^{\pm}$ for $j \in Q$.
Substituting into the decoupled form of 
${\Psi_r}_i^{(-)}$ in $R \ge R_0$ and after rearrangement, we obtain
${\Psi_r}_i^{(-)}$ = $\sum _{j \in P} {\tilde{\Psi}}_j^{(-)}
V_{ji}$ ($i \in P$). Likewise, we obtain
${\bm{\Psi_r}}_i^{(-)}$ =
$\sum _{j \in P} {\tilde{\bm{\Psi}}}_j^{(-)} V_{ji}$. 
$V_{ji}$ may be denoted in Dirac notation as 
either 
$\bigl(\tilde{\bm{\Psi}}_j^{(-)}|{\bm{\Psi_r}}_i^{(-)}\bigr)$
or
$\bigl(\tilde{\Psi}_j^{(-)}|{\Psi_r}_i^{(-)}\bigr)$, with the
precaution that it should not be interpreted as an integral.
Then showing that 
$\bigl(\tilde{\Psi}_j^{(-)}|{\Psi_r}_i^{(-)}\bigr)$ =
$\bigl(\tilde{M}_j^{(-)}|{\Psi_r}_i \bigr)$
is  equivalent to showing that ${\Psi_r}_i$ 
=
$\sum_{j \in P} \tilde{M}_j^{(-)} V_{ji}$  where $i \in P$.
 
The proof hinges on the following relation:
\begin{equation}
\tilde{M}_j^{(-)} = \tilde{\Psi}_j^{(-)} + 
i \tilde{\xi} \tilde{\Psi}_c^{(-)} V_{1j}^{\dag} .
\label{M_k}
\end{equation}
Let us derive the relation. The 
coefficient of $\tilde{\Psi}_c^{(-)}$ of 
Eq. (\ref{tilde_M_N}), i.e.,  
$\bigl[ \tilde{K}^{co} 
(-i+\tilde{K}^{oo})^{-1} \bigr]_{ck} $, can be recast as 
$- (\tilde{S}^{cc}+1)^{-1} \tilde{S}^{co}$. From $\tilde{S}^{cc}$ = 
$S_r^{cc}$, $\tilde{S}^{co}$ = $S_r^{co} V^{\dag}$ and 
$(S_r^{cc}+1)^{-1} S_r^{co}$ = $-i \tilde{\xi} (1,0,...,0)$,
Eq. (\ref{M_k}) is easily obtained.
Now, from the relation
${\Psi_r}_i$ = 
$\sum_{j \in P,Q} {\Psi_r}_j^{(-)} (1+i K_r )_{ji}$ 
between the standing-waves and incoming-waves 
with $i \in P$ hereinafter, 
one obtains
\begin{equation}
{\Psi_r}_i
= {\Psi_r}_i^{(-)} + i \tilde{\xi} {\Psi_r}_c^{(-)} \delta_{1i} .
\label{Psi_r_j}
\end{equation}
Substituting Eq. (\ref{M_k}) into
${\Psi_r}_i^{(-)}$ = $\sum_{j \in P} 
{\tilde{\Psi}}_j^{(-)} V_{ji}$, 
${\Psi_r}_i^{(-)}$ can be expressed as
$\sum_{j \in P} \tilde{M}_j^{(-)} V_{ji}$
$-$ $i\tilde{\xi} {\tilde{\Psi}}_c^{(-)} \delta_{1i}$,
from which one finally obtains
${\Psi_r}_i$ = $\sum_{j \in P} {\tilde{M}}_j^{(-)} V_{ji}$. 
Comparison of this with 
${\Psi_r}_i^{(-)}$ = $\sum_{j \in P} 
{\tilde{\Psi}}_j^{(-)} V_{ji}$
proves that 
$\bigl(\tilde{\Psi}_j^{(-)}|{\Psi_r}_i^{(-)}\bigr)$ equals
$\bigl(\tilde{M}_j^{(-)}|{\Psi_r}_i \bigr)$.
Notice that the right-hand side of Eq. (\ref{Psi_r_j}) 
is equal to 
${M_r}_i^{(-)}$ 
so that ${\Psi_r}_i $ = ${M_r}_i^{(-)}$. 
Since the projection factor 
$\bigl(\tilde{M}_j^{(-)}|{\Psi_r}_i \bigr)$
corresponds to $(\psi_E^{-(j)}|\psi_E^{(a)} )$ of CM,
the equality $\bigl(\tilde{M}_j^{(-)}|{\Psi_r}_i \bigr)$ =
$\bigl(\tilde{M}_j^{(-)}|{M_r}_i^{(-)} \bigr)$ 
emphasizes that $M^{(-)}$ functions correspond to the background
wavefunction shorn of the configuration mixing with a discrete 
state.



\begin{thebibliography}{27}
\bibitem[{Com()}]{Comment}
\bibinfo{note}{Orthogonal transformation besides phase renormalization is
  already implied in Cooke and Cromer's work as Ueda indicated in Ref.
  \cite{Ueda87}. See the paragraph around Eq. (58) of Ref. \cite{Cooke85}}.
\bibitem[{S_e()}]{S_eta}
\bibinfo{note}{Notice that $\bm{S}$ defined here is different from the usual
  physical scattering matrix in that it is defined with respect to
  $\phi_j^{\pm}$ instead of the usual $\exp (\pm i k_j R )$. This difference
  has no influence on the current work. See Ref. \cite{Lee02_a} for the
  detailed study of its influence.}
\bibitem[{Con()}]{Convention}
\bibinfo{note}{We take the convention that $\delta_r$ increases from zero as
  $\tilde{\beta}$ increases from $-\pi /2$. This convention implies that $\sin
  \delta_r \cos \tilde{\beta} > 0$ or $\cos \delta_r \sin \tilde{\beta} <0$}.
\end{thebibliography}

\end{document}